\documentclass[prb,preprint]{revtex4}

\usepackage{epsf}
\usepackage{epsfig}
\usepackage{textcomp}
\usepackage{dcolumn}
\usepackage{bm}

\begin{document}


\title{Exact-Exchange Kohn-Sham formalism applied to one-dimensional
periodic electronic systems}

\author{Stefan Rohra$^1$, Eberhard Engel$^2$ and Andreas G\"orling$^1$}
\affiliation{$^1$Lehrstuhl f\"ur Theoretische Chemie,
Universit\"at Erlangen-N\"urnberg, Egerlandstr. 3, D-91058 Erlangen, Germany}
\affiliation{$^2$Institut f\"ur Theoretische Physik,
J.W.Goethe-Universit\"at Frankfurt, 
Max-von-Laue-Str.\ 1, D-60438 Frankfurt/Main, Germany}

\date{\today}

\begin{abstract}
The Exact-Exchange (EXX) Kohn-Sham formalism, 
which treats exchange interactions
exactly within density-functional theory, is applied to one-dimensional
periodic systems. 
The underlying implementation does not rely on specific symmetries of the 
considered system and can be applied to any kind of periodic structure in 
one to three dimensions.
As a test system, $trans$-polyacetylene, both in form
of an isolated chain and in the bulk geometry has been investigated. 
Within the EXX 
scheme, bandstructures and independent particle response functions are
calculated and compared to experimental data as well as to data calculated
by several other methods. Compared to results from the local-density
approximation, the EXX method leads to an
increased value for the band gap, in line with similar observations for 
three-dimensional semiconductors. An inclusion of correlation potentials
within the local density approximation or generalized gradient
approximations leads to only negligible effects in the bandstructure.
The EXX band gaps are in good agreement with experimental data for bulk
$trans$-polyacetylene.
Packing effects of the chains in bulk $trans$-polyacetylene are found
to lower the band gap by about 0.5 eV.
\end{abstract}


\maketitle

\def\angst{\,\text{\AA}}

\newcommand{\elmat}[3]{\langle {#1} | {#2} |{#3} \rangle}

\def\RR{({\bf r})}
\def\RB{{\bf r}}       
\def\RP{({\bf r}')}
\def\RPP{({\bf r}'')}
\def\phiG#1{\phi^{\Gamma\!,\gamma}_{#1}}
\def\phiGT#1{{\phi^{\Gamma\!,\gamma}_{#1}}^T\!\!}
\def\phiU#1{\phi^{\Upsilon\!,\upsilon}_{#1}}
\def\phiGP#1{\phi^{\Gamma'\!,\gamma'}_{#1}}
\def\phiGPT#1{{\phi^{\Gamma'\!,\gamma'}_{#1}}^T\!\!}
\def\phiGM#1{\phi^{\Gamma_M,\gamma}_{#1}}
\def\phiGMT#1{{\phi^{\Gamma_M,\gamma}_{#1}}^T\!\!}
\def\phiGm#1{\phi^{\Gamma\!,\mu}_{#1}}
\def\phiGmT#1{{\phi^{\Gamma\!,\mu}_{#1}}^T\!\!}
\def\nG#1{n^{\Gamma}_{#1}}
\def\nGP#1{n^{\Gamma'}_{#1}}
\def\nGM#1{n^{\Gamma_M}_{#1}}
\def\phiL#1{\phi^{\Lambda,\lambda}_{#1}}
\def\phiLT#1{{\phi^{\Lambda,\lambda}_{#1}}^T\!\!}
\def\nL#1{n^{\Lambda}_{#1}}
\def\gG{g^{\Gamma}}
\def\gGP{g^{\Gamma'}}
\def\gGM{g^{\Gamma_M}}
\def\gL{g^{\Lambda}}
\def\eG#1{\varepsilon^{\Gamma}_{#1}}

\def\coulP{|{\bf r}'-{\bf r}''|}
%

\section{\label{sec:introduction}Introduction}

The bandstructure is the key quantity for studying the 
electronic structure of solids and of periodic systems in general. 
Presently, in most cases, Kohn-Sham (KS) bandstructures, calculated 
either within the local density approximation (LDA) 
\cite{yang89,dreizler90} or within some generalized 
gradient approximation (GGA) \cite{yang89,dreizler90,koch00},
represent the starting point for such investigations.
In particular, the GW method
\cite{hedin65,hybertsen85,aryasetiawan98,aulbur00,onida02}, the currently most 
popular approach for the calculation of quasi-particle bandstructures 
and band gaps,
is usually based on LDA or GGA bandstructures. 
While it is possible to perform fully self-consistent GW calculations, 
this is usually not done, because such calculations are not only 
computationally more demanding than the non-self-consistent treatment 
based on LDA or GGA one-particle states and eigenvalues, but
also in the few cases where such calculations have been carried out for solids
worse results compared to experiment \cite{schoene98} were obtained. 
While the non-self-consistent GW approach based on KS input is somewhat 
unsatisfactory from a formal point of view, its success nevertheless
underlines the central role of the KS bandstructure. 
Moreover, also the methods to investigate optical properties usually start
from the KS bandstructure, as for instance the Bethe-Salpeter method 
\cite{onida02,hanke79,albrecht98,benedict98,rohlfing98}
or methods based on time-dependent density-functional theory
\cite{runge84,gross96,casida95,gorling98b,gorling99d,kim02}
or time-dependent current-density-functional theory
\cite{berger05}.

Despite their central role for the investigation of electronic systems
LDA and GGA bandstructures suffer from severe shortcomings. 
The most important of these is the substantial underestimation of the 
band gaps of semiconductors and insulators 
\cite{filippi94,ortiz92,perdew83,sham83,sham85,perdew86}.
This fact was often attributed to the KS formalism itself.
Indeed, even the exact KS band gap, i.e.,\ the gap which would result 
from the exact KS exchange-correlaton potential, is not identical
with the physical band gap \cite{perdew83,sham83}. 
The two gaps differ by the derivative discontinuity
of the exchange-correlation potential at integer electron numbers
\cite{perdew83,sham83}.
As a matter of fact, KS orbitals and eigenvalues originally were considered 
as auxiliary quantities with no or little physical meaning. Recently, however,
it was shown that KS eigenvalues and their differences are physically 
meaningful and, e.g., 
are well-defined zeroth order approximations 
for ionization and excitation energies
\cite{perdew82,almbladh85,chong02,gritsenko02,goerling96,filippi97}
as well as for band gaps \cite{goerling95}. In order to clarify the meaning of
these zeroth order approximations it is instructive to consider how excitation
energies and band gaps are given in density-functional theory
(DFT). Excitation energies result from time-dependent DFT. For excitation
energies that refer to excitations from an 
occupied to an unoccupied orbital within a one-particle picture, 
TDDFT excitation energies can be considered
as the sum of the corresponding eigenvalue difference plus a TDDFT
correction. Band gaps, on the other hand, are given by the smallest
eigenvalue difference between occupied and unoccupied orbitals, i.e., the KS
band gap, plus the derivative discontinuity. At the level of the zeroth order
approximation of KS eigenvalue differences, TDDFT corrections or
derivative discontinuities, respectively, are neglected. Therefore optical
absorption onsets, i.e., optical
band gaps, related to excitation energies and quasiparticle band gaps
including fundamental band gaps are not distinguished at the level of KS
eigenvalues 
differences. Nevertheless, eigenvalue difference can be reasonable
approximations for quasiparticle band gaps and excitation energies 
as discussed below.

The shortcomings of LDA and GGA 
bandstructures, at least in most cases, seem to predominantely originate from
the exchange-correlation functionals used,
rather than from limitations of the KS formalism itself. The major problem of
the LDA and GGA is the incomplete cancellation of the unphysical 
self-interaction of each electron with itself, which is contained 
in the Coulomb energy and potential. 
These Coulomb self-interactions raise the KS eigenvalues. 
The more localized valence bands are affected more strongly than the 
less localized conduction bands.
As a result of the different impact of the self-interaction error on
valence and conduction bands the LDA or GGA band gaps are artificially 
decreased. 
While, in principle, the self-interaction contained in the 
Coulomb potential is exactly cancelled by the KS exchange potential, 
the LDA and GGA for the latter not nearly lead to a complete cancellation.

Recently, a KS approach for solids 
treating the exchange energy and potential exactly 
was introduced \cite{kotani95,gorling96,stadele97,stadele99}. 
This exact-exchange (EXX) approach does not suffer
from the spurious self-interaction and yields band gaps that are
much larger than those from LDA or GGA calculations. For standard 
semiconductors band gaps in excellent agreement 
with experiment were obtained 
\cite{stadele97,stadele99,qteish05,rinke05,sharma05}.  
In these calculations the correlation potential was either completely
neglected or approximated within the LDA or GGAs. It turned out that the
choice of the treatment of the correlation potential has little effect on the
bandstructure. The agreement between experimental and EXX band gaps for 
standard semiconductors shows that for these materials the approximations
made in the EXX approach, the neglect of the derivative discontinuity and the
approximate treatment of the correlation potential together have little effect.
For these materials the EXX approach therefore is an alternative to the
GW method. The EXX approach, from a formal point of view, has
the advantage that it is a pure KS procedure, which is carried out 
self-consistently. Furthermore, the more physical
EXX band structures can be used instead of LDA or GGA bandstructures as
basis of GW \cite{aulbur00b,fleszar01}, Bethe-Salpeter,
or time-dependent density-functional methods \cite{kim02}.

Note that EXX band structures, even in the exchange-only case, are completely
different from Hartree-Fock (HF) band structures. Both exact exchange-only KS
and HF methods treat exchange exactly, however, the underlying exchange
energy and potential are defined differently. The exchange energy in both cases
is the exchange energy of the corresponding Slater determinant, the HF
determinant or the KS determinant, respectively. However, the latter two
determinants, and therefore also the orbitals building them, are different and
subsequently also the HF and exchange-only KS exchange energies are
different. The HF
and KS orbitals and thus the corresponding determinants are different because
the HF and exchange-only KS one-particle equations contain a different
exchange potential. The KS exchange potential is defined as the functional
derivative of the exchange energy with respect to the electron density and
therefore, by definition, is a local multiplicative potential. The local
multiplicative KS exchange potential acts in the same way on all orbitals, be
they occupied or unoccupied. The exact exchange potential therefore cancels
the unphysical Coulomb self-interactions contained in the Coulomb potential for
all orbitals. The nonlocal HF exchange potential acts in a different way on
occupied and unoccupied orbitals and as a result only the occupied orbitals are
unaffected by Coulomb self-interactions and describe electrons in the
system under consideration. The unoccupied orbitals, on the other hand,
describe electrons in the corresponding $N+1$-particle system with $N$ being
the number of electrons. As a result HF band structures usually exhibit a
much too large band gap and in case of metals a singularity at the Fermi
level. Exact exchange-only KS bandstructures, on the other hand, are not
affected by such problems.

While EXX methods yield very good results for standard semiconductors, 
an EXX study of solid noble gases leads to a somewhat mixed picture
\cite{magyar04}. 
The calculated band gaps are clearly superior to the LDA values, 
but are still significantly below the experimental data. 
So far, EXX methods were only applied to three-dimensional periodic
systems (as well as to atoms \cite{talman76} and molecules
\cite{gorling99,ivanov99}). 
It seems desirable to investigate the performance of EXX methods also 
for other types of periodic systems. 
In this work the EXX approach is applied to the electronic structure 
of $trans$-polyacetylene, as a typical representative of an organic 
one-dimensional periodic system. One-dimensional
organic oligomeres or polymeres are of great interest due to their 
high potential as active materials in new optoelectronic devices.
 
$Trans$-polyacetylene is the prototype example for
an organic one-dimensional periodic system. While an isolated
$trans$-polyacetylene chain may be considered as a molecular wire or
one-dimensional semiconductor, in practice usually fibers of
$trans$-polyacetylene or crystalline bulk $trans$-polyacetylene are
encountered \cite{kahlert,fincherprl}.
Despite the fact that crystalline $trans$-polyacetylene is a three-dimensional
periodic material its properties are mostly determined by the periodicity along
the polyacetylene chain. We therefore occasionally address all forms of
$trans$-polyacetylene as one-dimensional periodic materials in this work
despite the fact that it is shown below that there exists some nonnegligible
packing effects in crystalline $trans$-polyacetylene. Because
$trans$-polyacetylene is a prototypical one-dimensional periodic organic
system it is 
not surprising that a
number of theoretical studies of the optical
properties of this system have been carried out 
\cite{liegener88,suhai92,sun96,foerner97,rohlfing99,ayala01,birkenheuer04}.
In Ref. \onlinecite{rohlfing99} the band gap and the 
optical spectrum of an isolated  
$trans$-polyacetylene chain were calculated with the 
GW and the Bethe-Salpeter method, respectively. The results, e.g.,\ the
band gap of 2.1 eV and the absorption spectrum with a single peak at
1.7 eV, agree well
with experimental data \cite{fincher79,leising88}, 
which, however, refer to bulk
$trans$-polyacetylene. In Ref. \onlinecite{birkenheuer04}, on the other hand, 
a much larger band gap 
of 4.1 eV for an isolated  $trans$-polyacetylene chain was obtained by 
a multireference configuration interaction  approach.  
Finally, in Refs. \onlinecite{sun96} and \onlinecite{ayala01}, 
band gaps of 3.2 and 3.7 eV, 
respectively, emerged from 
M$\o$ller-Plesset perturbation theory for an 
isolated $trans$-polyacetylene chain.
As most of these studies considered isolated chains, but relied on 
experimental data from bulk polyacetylene for comparisons, it seems 
desirable to investigate the influence of the geometry and the 
arrangement of polyacetylene chains on the bandstructure.
Such an investigation is the second goal of this work, besides the 
assessment of the performance of the EXX approach for one-dimensional 
periodic organic systems.

\section{\label{sec:method} Method}
%
%
%
Within the EXX KS approach 
\cite{kotani95,gorling96,stadele97,stadele99,sharp53,talman76,KLI90a,EV93a,gorling94,gorling95b},
the exchange energy as well as the local 
multiplicative KS exchange potential, not to be confused with the 
nonlocal Hartree-Fock exchange potential, are treated exactly, rather
than via some approximate density-functional.
For a periodic system the exact exchange energy per unit cell is given by
\begin{equation}
E_x ({\bf r}) = - (1/N_{\bf k}) \, \sum_{vv^\prime {\bf kk^\prime}}
\int \int d {\bf r} d {\bf r^\prime}
\frac{\phi^\ast_{v{\bf k}}({\bf r})
\phi_{v^\prime{\bf k^\prime}}({\bf r})
\phi^\ast_{v^\prime{\bf k^\prime}}({\bf r^\prime})
\phi_{v{\bf k}}({\bf r^\prime})
} 
{|{\bf r} - {\bf r^\prime}|}.
\label{exenergy}
\end{equation}
In Eq.\ (\ref{exenergy}), $\phi_{v{\bf k}}$ are valence 
orbitals, with crystal momentum ${\bf k}$. 
Orbitals generally are assumed to be
normalized with respect to the crystal volume. In the considered case of
non-spin-polarized systems, the occupied orbitals are doubly
occupied and spin is taken into account by appropriate prefactors of 2.
The summation runs over all valence bands, $v$ and $v'$, and over all 
${\bf k}$-points, ${\bf k}$ and ${\bf k}'$ within the first Brillouin
zone. The number of ${\bf k}$-points is denoted by $N_{\bf k}$

The KS exchange potential $v_x$ is defined as the functional derivative 
$v_{x}({\bf r}) = {\delta E_x [\rho]}/{\delta \rho ({\bf r})}$ of the
exchange energy with respect to the electron density. It obeys the equation
\begin{equation}
\int \! d{\bf r^\prime} \, X_s ({\bf r},{\bf r^\prime}) 
\, v_{x}({\bf r^\prime}) 
= 2 \sum_{vc{\bf k}} \,
\Big[ \langle v {\bf k} | \hat{v}_x^{NL} | c{\bf k} \rangle \,
\frac{\phi^\ast_{c{\bf k}}({\bf r}) \phi_{v{\bf k}}({\bf r)}}
{\epsilon_{v {\bf k}} - \epsilon_{c {\bf k}}} + \mathrm{c.c.} \Big] \,,
\label{exx}
\end{equation}
which shall be called EXX equation. The $\phi_{c{\bf k}}$ occuring on the
right hand side of the EXX equation are unoccupied orbitals forming
conduction bands $c$ and 
%
the nonlocal operator $\hat{v}_x^{NL}$ 
with the kernel
\begin{equation}
\hat{v}_x^{NL} ({\bf r},{\bf r^\prime}) = - \sum_{v {\bf q}}
\frac{\phi_{v{\bf q}}({\bf r}) \phi^\ast_{v{\bf q}}({\bf r^\prime})}
{|{\bf r} - {\bf r^\prime}|}
\end{equation}
is a non-local exchange 
operator of the form of the Hartree-Fock
exchange operator, but built from KS orbitals. The eigenvalues of the orbitals
$\phi_{v{\bf k}}$ and $\phi_{c{\bf k}}$ are denoted by  
$\epsilon_{v {\bf k}}$ and $\epsilon_{c {\bf k}}$, respectively.
The independent particle response function $X_s$, i.e., the response 
function of the KS system, is given by
\begin{equation}
X_s ({\bf r},{\bf r^\prime}) =
\frac{\delta \rho ({\bf r})}{\delta v_{s} ({\bf r^\prime})} =
2 \sum_{vc{\bf k}} \, \frac{\phi^\ast_{v{\bf k}}({\bf r})
\phi_{c{\bf k}}({\bf r})
\phi^\ast_{c{\bf k}}({\bf r^\prime})
\phi_{v{\bf k}}({\bf r^\prime})}
{\epsilon_{v {\bf k}} - \epsilon_{c {\bf k}}} + \mathrm{c.c.}
\label{chi0}
\end{equation}
In Eq. (\ref{chi0}), $v_{s}$ is the effective KS potential.

If the exchange potential $v_{x}$, the response function $X_s$, and the right
hand side of the EXX equation (\ref{exx}) is expanded in an auxiliary basis
set, here an auxiliary plane wave basis set, then the EXX equation turns into
a matrix equation, which can easily be handled computationally
\cite{gorling96,stadele97,stadele99}.

\section{\label{sec:compdetails} Computational Details}

Originally  the EXX formalism was implemented in a plane-wave
code optimized for
three-dimensional periodic cubic systems \cite{stadele99} by exploiting the
cubic symmetry and by using the concept of special $\bf{k}$-points
\cite{monkhorst76,froyen89}. 
For the investigation of non-cubic systems, including one- or two-dimensional
periodic systems via the 
super cell ansatz, we programmed 
a new plane-wave implementation, which does not rely on symmetry or 
special $\bf{k}$-points. In super cell calculations the directions along the
artificial periodicity due to the super cell ansatz are treated like the other
direction related to real physical periodicity in the sense that no other
plane wave cutoffs or artificial screenings are introduced. Merely, the number
of $\bf{k}$-points is set to one for those directions in the Brillioun zone
that correspond to the directions along the
artificial periodicity. In order to test that the super cells were
chosen large enough we additionally carried out calculations with more than one
$\bf{k}$-point along the directions corresponding to the artificial
periodicity and checked that results did not change and, in particular, that no
significant dispersion occured in those directions.

The
cutoff for the plane wave basis set to represent the orbitals 
and the cutoff for the auxiliary plane wave 
basis set representing the KS response 
function $X_s$ and the exchange potential $v_x$ in all cases were chosen 
to be 32 Ry and 12 Ry, respectively. While these cutoffs are not sufficient for
calculations of total energies, they turned out to be sufficient for the 
calculation of bandstructures. Errors in the resulting 
band gaps can be estimated to be below the order of about 0.05 eV, 
which is more than sufficient for the purpose of this work.

Instead of special $\bf{k}$-points,
we used uniform 
grids of $\bf{k}$-points covering the first Brillouin zone. 
For selfconsistent KS calculations of isolated chains of polyacetylene we
used 24 $\bf{k}$-points 
in the direction of the chain, which ensured very good convergence with the
number of $\bf{k}$-points. 
For selfconsistent KS calculations of bulk $trans$-polyacetylene additional
$\bf{k}$-points  were placed in the
two directions along the reciprocal unit cell vectors not corresponding to
the chain direction. Here, a 
uniform 2$\times$32$\times$2
mesh (32 $\bf{k}$-points along the chain direction, 2 $\bf{k}$-points along
the two other directions) was employed. The calculations of band structures
and KS response functions, as usually, are based on the KS potentials obtained
in the selfconsistent calculations, i.e., the band structures and KS response
functions were calculated by diagonalizing, for the required $\bf{k}$-points, 
KS Hamiltonian operators containing the KS potentials from the selfconsistent
calculations. (Note that the exact KS exchange potential, in contrast to the
HF exchange potential, does not depend on the crystal momentum $\bf{k}$ of the
orbital it acts upon.) For the KS response functions of isolated chains of
polyacetylene we used a uniform 1$\times$350$\times$1 mesh of $\bf{k}$-points,
for bulk $trans$-polyacetylene a uniform 5$\times$150$\times$5 mesh was
employed.

In all calculations all unoccupied states were
taken into account for both
the construction of the response function and the right hand side of
the EXX equation.
Typical values for the number of conduction bands were 3600 (1500)
for the isolated chain (bulk) case.


In the case of EXX calculations the use of pseudopotentials is
particularly attractive in view of the large number of Fock matrix
elements to be evaluated in all-electron calculations.
For that reason EXX pseudopotentials have been introduced quite
early
\cite{BK95,stadele97,stadele99,moukara00}.
However, it has been noticed immediately that a spurious long-range
structure develops in the EXX pseudopotential
\cite{BK95,stadele97,stadele99,moukara00}, if
the standard procedure for the construction of normconserving 
pseudopotentials \cite{HSC79,H89,TM91} is utilized.
This long-range structure, which effectively simulates an additional
artificial charge on the ion, can be traced to the core-valence
interaction \cite{EHSDC01}.
In order to eliminate this structure in a systematic fashion a
self-consistent pseudopotential construction scheme has been 
introduced \cite{EHSDC01}, in which a clean asymptotic behavior
is enforced as constraint.
The resulting EXX pseudopotentials have been shown to give very 
accurate atomic excitation energies and spectroscopic constants
of molecules.
In Ref.\ \onlinecite{EHSDC01} it has also been demonstrated explicitly that, 
apart from the long-range structure eliminated via the self-consistent
pseudopotential construction, the core-valence interaction is less 
relevant in the case of the exact exchange than for the LDA.
\par
In the present work, all calculations with the exact exchange were
based on normconserving pseudopotentials obtained by this 
self-consistent pseudopotential construction scheme.
In all cases the density-functionals (LDA, exact exchange-only, exact 
exchange plus approximate correlation) used for the generation of the 
pseudopotentials and in the self-consistent groundstate calculations 
of the periodic system were chosen to be identical (pure LDA 
pseudopotentials were generated by the standard procedure \cite{TM91}).
For hydrogen the pseudopotentials contained exclusively an angular 
momentum component with $l=0$.
For carbon pseudopotentials with $s$- and $p$-components were employed, 
with the $s$-component chosen as local component for all calculations. 
Additional tests with the $p$-component as local component for the 
carbon atom were carried out, but did not lead to significant changes 
of the results. 
For the cutoff radii values of $r_{c,l=0}^H=0.9$, $r_{c,l=0}^C=1.2$ 
and $r_{c,l=1}^C=1.1$ (in units of Bohr radii) were used, following
Ref.\cite{EHSDC01}.

\section{\label{sec:results} 
Results}

\subsection{\label{sec:isolatedchain} Isolated, infinite chain of 
$trans$-polyacetylene}
As a first system we consider an 
isolated, infinite chain 
of $trans$-polyacetylene, relying on the super cell concept. 
The polymere is planar and exhibits alternating single and double 
bonds between the carbon atoms, which are accompanied by different 
bond lengths. The unit cell contains a $\mathrm{C_2 H_2}$ unit.
The bond distances of the carbon 
backbone were taken from experimental 
data from X-ray scattering of crystalline 
$trans$-polyacetylene \cite{fincherprl}. This geometry, which shall be named
G1, has also been used in a recent
investigation of electronic correlation effects in $trans$-polyacetylene
\cite{birkenheuer04}. The carbon-carbon bond lengths in geometry G1 
are given by 
$d(\mathrm{C-C})=1.45\mathrm \AA$ and $d(\mathrm{C=C})=1.36\mathrm \AA$. 
The lattice constant along the chain direction equals $2.455\mathrm \AA$
\cite{kahlert}. A ($\mathrm{C-H}$)--bond length of $1.087\mathrm \AA$
has been found by optimization on the HF and MP2 level \cite{suhai92}.
The angle $\angle(\mathrm{C-C,C=C})$ (with respect to the short
($\mathrm{C=C}$)-- double bond) has been determined to be $120.057$\textdegree
via HF optimization \cite{birkenheuer04}.
\\
The system was placed in the 
the $xy$-plane with the chain oriented along the $y$-direction. The
unit cell is defined by the vectors
${\bf a}_1=(a,0,0)$, 
${\bf{a}_2}=(0,c,0)$ and ${\bf{a}_3}=(a/2,0,b)$ with the parameter 
$c=2.455\mathrm \AA$
defining the lattice vector along the polyacetylene chain. The parameter $a$
was set to $a=8\mathrm \AA$. The third lattice vector was chosen 
in a way that the 
$p$-orbitals of the carbon atoms of two adjacent chains 
do not lie on top of each other, but at maximal distance. The parameter $b$ was
set to $b=9\mathrm \AA$. The values of $a=8\mathrm \AA$ 
and $b=9\mathrm \AA$ turned out to be large enough that 
the chains can be considered as isolated. Effects on the band gap 
due to coupling of the chains
can be estimated by variation of the super cell size to be below 0.05 eV.
For further details of the geometry see Appendix.

\begin{figure}[h]
\includegraphics*[width=6.5cm]{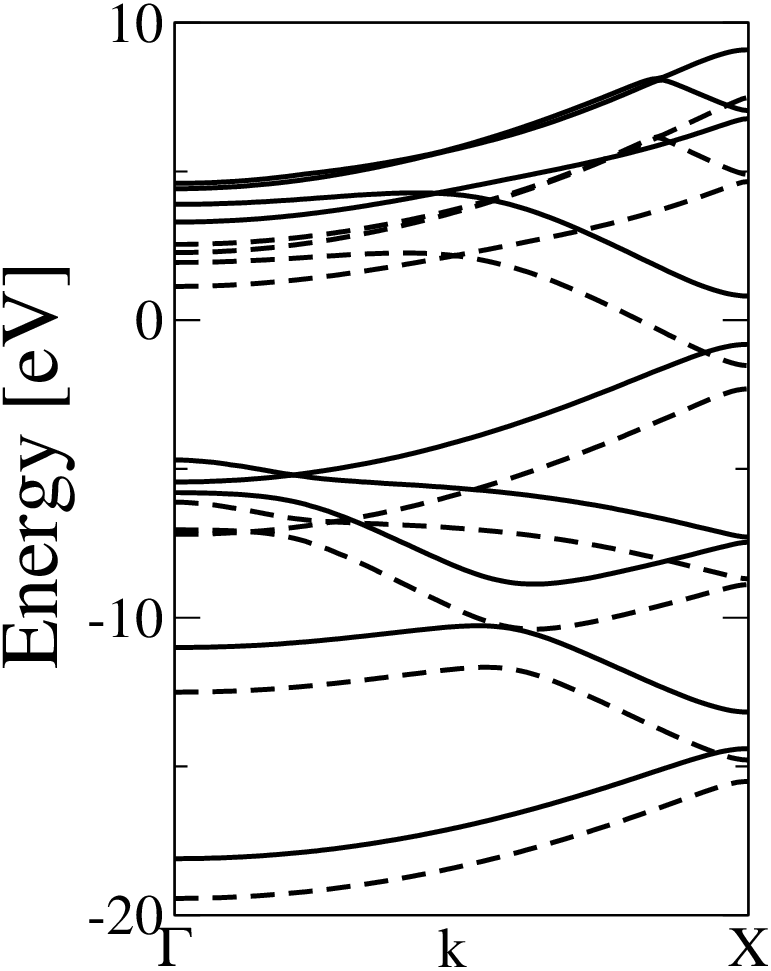}
\caption
{Bandstructure for an isolated, infinite chain of polyacetylene for the 
LDA (dashed lines) and the EXX (solid lines) case}
\label{baiso}
\end{figure}

The calculated LDA band gap equals 0.78 eV, whereas the EXX calculation
yields a band gap of 1.63 eV.
The bandstructure for these two cases is displayed in Fig.\ \ref{baiso},
adjusted with respect to each other by placing the Fermi energy at 0 eV. 
The LDA and EXX bandstructure are qualitatively similar, however, 
the energetical difference between valence and conduction bands is
significantly smaller in the case of the LDA, leading to the 
observed small gap.
This is in line with the finding that for three-dimensional semiconductors 
the EXX approach increases the commonly
too small LDA band gap. Addition of an LDA correlation potential to the EXX
potential leads to a band gap of 1.65 eV, i.e., increases the gap marginally 
by 0.01 eV, whereas addition of a GGA-correlation potential, here
the PBE correlation potential, to the EXX potential marginally decreases 
the gap to 1.61 eV.  This shows, again in line with the observation for
three-dimensional semiconductors, that the treatment of  the 
correlation potential has no significant influence
on the band gap. Within the EXX scheme, the band gap 
comes close to the 
value of 2.1 eV resulting from GW calculations \cite{rohlfing99}, 
but still stays somewhat smaller. 
The calculated  band gaps as well as those from several other methods are
collected in Table \ref{bandgaps}.

To investigate the sensitivity of the bandstructure of an isolated
polyacetylene chain on the geometry, we calculated 
the EXX bandstructure also in a slightly modified geometry, denoted G2, 
which has also been used in Ref. \onlinecite{rohlfing99}.
The geometry G2 is characterized by the carbon-carbon bond lengths 
$d(\mathrm{C-C})=1.44\mathrm \AA$ and 
$d(\mathrm{C=C})=1.36\mathrm \AA$ \cite{yannoni}. Optimization yielded
\cite{rohlfing99} a lattice constant
along the chain direction of $2.473\mathrm \AA$.
A carbon hydrogen bond length
$d(\mathrm{C=H})=1.1\mathrm \AA$ \cite{rohlfing99} and an angle
$\angle(\mathrm{C-C,C=C})$ of 118\textdegree have been used (see also
Appendix).
In geometry G2 an EXX band gap of 1.56 eV is obtained, which differs 
only slightly from the result observed for the geometry G1.


\subsection{\label{sec:bulk} Bulk $trans$-polyacetylene}

The geometries of the chains building the considered bulk 
$trans$-polyacetylene 
are those of the above discussed isolated chains in geometry G1, 
i.e., the geometry based on X-ray structure \cite{kahlert}.
Within the crystalline
$trans$-polyacetylene, each unit cell contains two polyacetylene chains. 
The unit cell according to Ref.\ \onlinecite{fincherprl} 
corresponds to a simple monoclinic 
Bravais lattice with the values of $a=4.24\mathrm \AA$, $b=7.32\mathrm \AA$ 
and $c=2.46\mathrm \AA$ for the lattice constants (see Appendix for further
details).

\begin{figure}[h]
\includegraphics*[width=6.5cm]{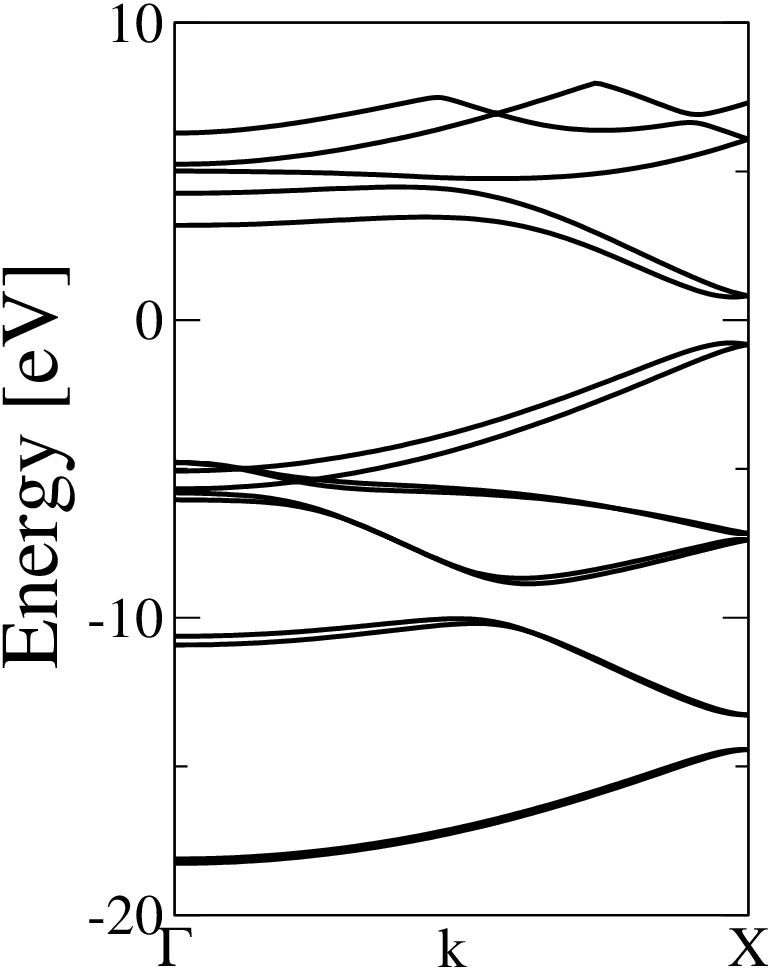}
\caption
{Bandstructure for bulk polyacetylene for the EXX case}
\label{babulkexx}
\end{figure}

\begin{figure}[h]
\includegraphics*[width=6.5cm]{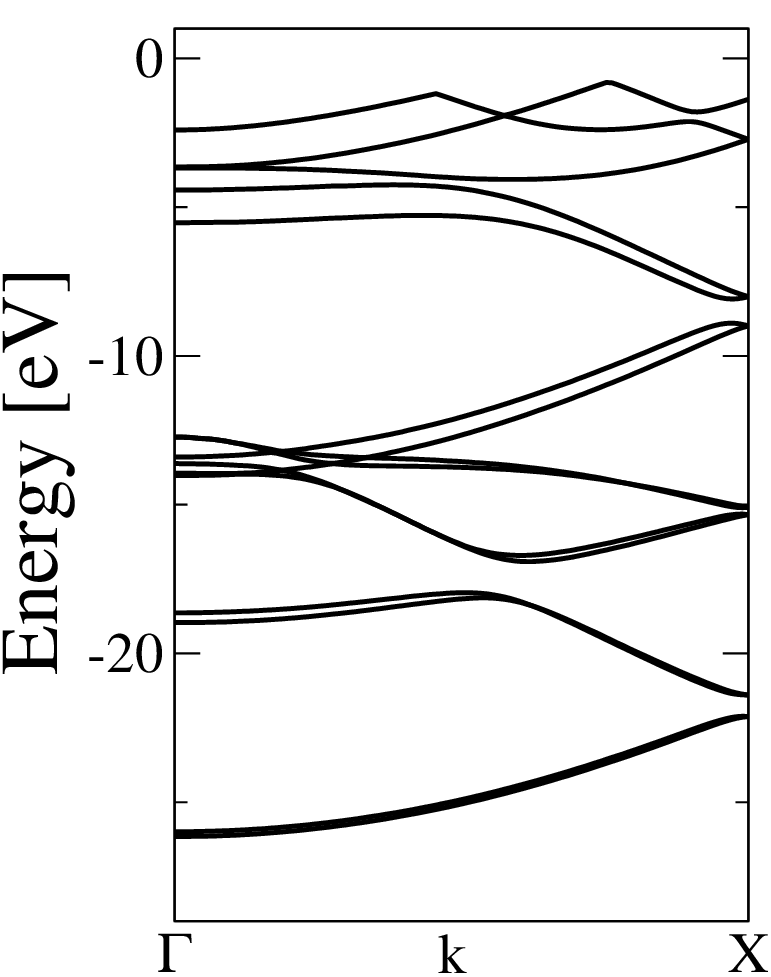}
\caption
{Bandstructure for bulk polyacetylene for the LDA case}
\label{babulklda}
\end{figure}

\begin{figure}[h]
\includegraphics*[width=6.5cm]{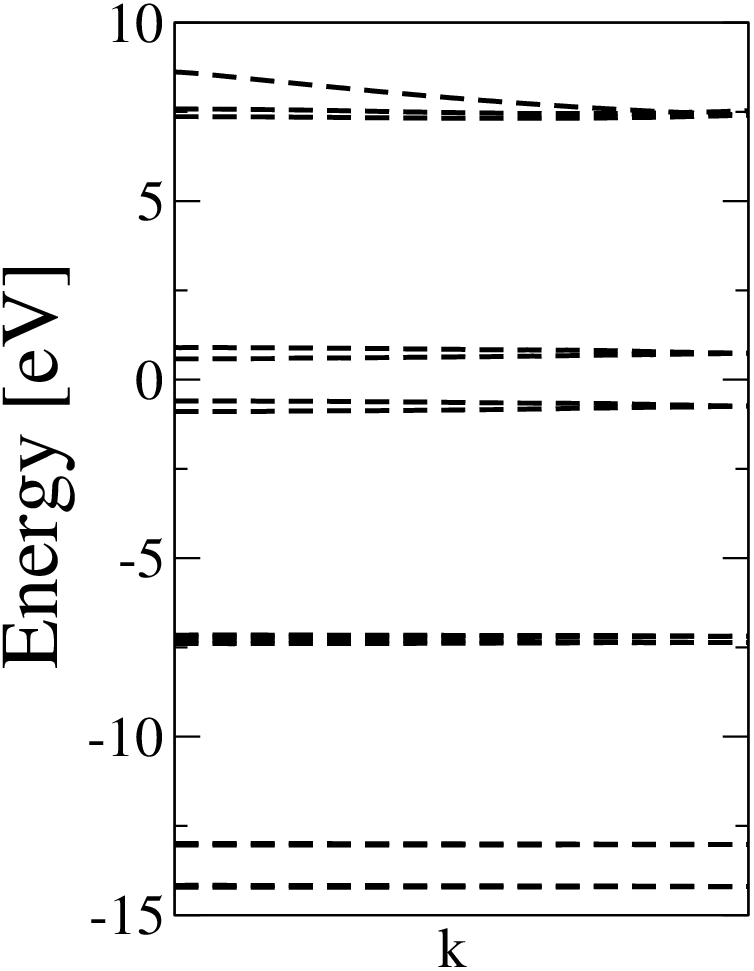}
\caption
{Dispersion of the band gap for bulk polyacetylene (EXX case)
along the line from point
(0.5,0.5,0.0) to point (0.5,0.5,0.5) in multiples of the reciprocal
lattice vectors}
\label{dispbulkexx}
\end{figure}

\begin{figure}[h]
\includegraphics*[width=6.5cm]{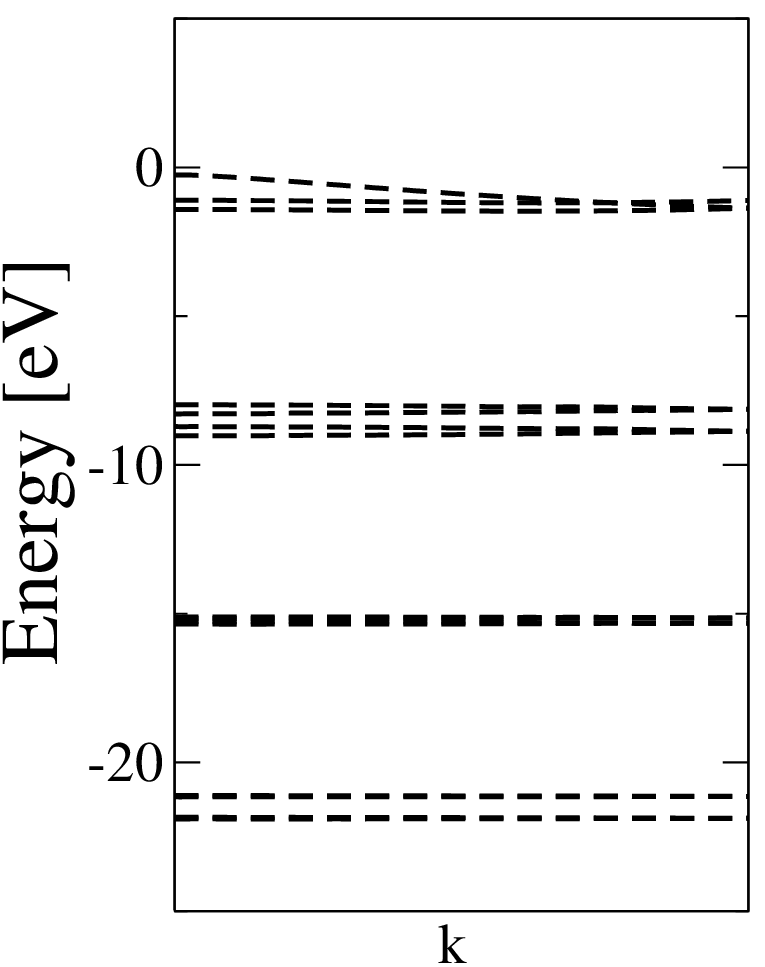}
\caption
{Dispersion of the band gap for bulk polyacetylene (LDA case)
along the line from point
(0.5,0.5,0.0) to point (0.5,0.5,0.5) in multiples of the reciprocal
lattice vectors}
\label{dispbulklda}
\end{figure}

Again we carried out LDA and EXX calculations, which yielded band gaps 
at the $X$-point of 0.98 eV 
and 1.64 eV. The corresponding bandstructures are displayed 
in Fig.\ \ref{babulkexx} and \ref{babulklda}. However in both cases the band
gap at the $X$-point is not the fundamental band gap. Along the line 
from the $\Gamma$- to the $X$-point, one observes band gaps
(0.81 eV for LDA and 1.55 eV for EXX), which are
reduced in comparison to the gaps at the $X$-point. However, this band gap
also is not the fundamental band gap. For LDA as well as EXX calculations,
the smallest band gaps we found are  
located at the point (0.5,0.5,0.0), in multiples of the reciprocal 
lattice vectors,
(see Appendix for numbering of reciprocal lattice vectors), i.e., at 
the edge of the Brillouin zone. At this point we calculated band gaps of
0.43 eV for LDA and 1.18 eV for EXX,
which represent, compared to the band gap at the $X$-point, 
a remarkable dispersion
of about 0.55 eV for LDA, resp. 0.46 eV for EXX. This finding is in agreement
with results from Hartree-Fock calculations which estimated a reduction of the
band gap by 0.6 eV due to interband interactions \cite{birkenheuer04}. 
The LDA and EXX bandstructures along the
line from point (0.5,0.5,0.0) to point (0.5,0.5,0.5) in multiples of the
reciprocal lattice vectors are shown in Fig.\ \ref{dispbulkexx} 
and \ref{dispbulklda}.

Furthermore, we also checked the influence of the addition of an LDA or PBE
correlation potential to the EXX potential. Like in the case of
an the isolated chain, the correlation potential turned out
to have an only  negligible effect on the bandstructure.

Optical absorption experiments for $trans$-polyacetylene in the considered
geometry 
\cite{fincher79} give an absorption coefficient
rising sharply at 1.4 eV and having a peak at about 1.9 eV. A
second similar
measurement \cite{leising88} gives an identical value for the peak and a value
of 1.5 eV for the onset of the absorption. The lattice constants mentioned
within the framework of this second measurement \cite{kahlert}, 
differ slightly 
from the geometry of Ref. \onlinecite{fincher79}, underlying
our calculations ($a=4.18\mathrm \AA$,
$b=7.34\mathrm \AA$, $c=2.455\mathrm \AA$).
Nevertheless, the experimental results for the two distinct measurements agree 
more or less and are
in reasonable agreement with the band gaps from our EXX calculations.
(Remember that no distinction between quasi particle
and optical band gaps is made at the level of a zeroth order approximation
relating  KS eigenvalue differences to observables.) 
The data of the band gaps for all discussed cases have been compiled in Table 
\ref{bulkbandgaps}.\\


\subsection{\label{sec:independent} Reponse functions}

\begin{figure}[h]
\includegraphics*[width=6.5cm]{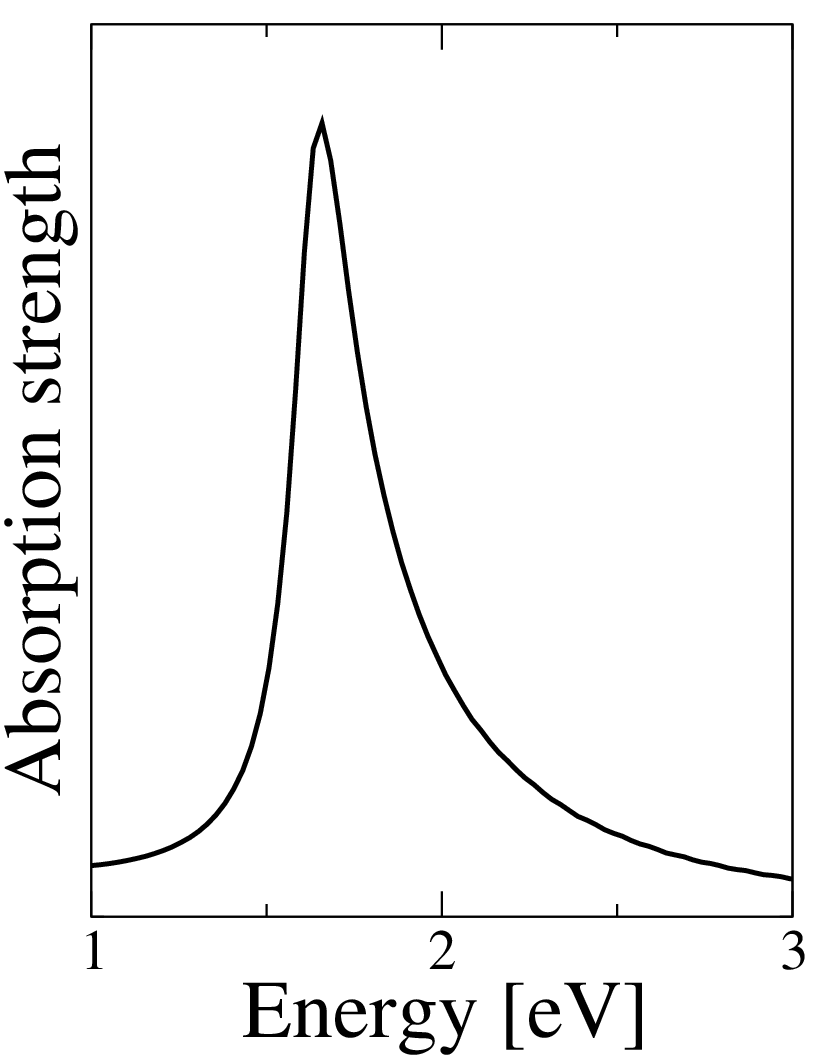}
\caption
{EXX Kohn-Sham response function for an isolated, infinite chain of 
polyacetylene}
\label{exxindep}
\end{figure}

The optical absorption strength obtained by the independent particle, i.e., 
Kohn-Sham, response function for an isolated $trans$-polyacetylene
chain in geometry G1 is shown
in Fig.\ \ref{exxindep} for the EXX-case. The onset of the
peak coincides with the calculated fundamental band gaps, 
the absorption maximum lies about 0.1 eV higher than the band gaps.
A similar behavior was obtained for the Kohn-Sham response function
for the LDA-case with the corresponding peak shifted to a lower
energy. 
The absorption spectrum resulting from the EXX response function 
is quite similar to the one obtained via the Bethe-Salpeter equation
\cite{rohlfing99}, which includes electron-hole
interactions, i.e., excitonic effects, that are missing in an independent
particle spectrum. This could be due to a
cancellation of two effects. For an isolated $trans$-polyacetylene chain
the neglect of electron-hole
interactions seems to shift \cite{rohlfing99} the peak of the absorption
spectrum to higher energy values. This could cancel the effect that the
EXX band gap is smaller than the one obtained within the GW scheme
\cite{rohlfing99}. In any case the similarity between the Bethe-Salpeter and
the EXX absorption spectra suggests that, beyond a possible shift of the
spectrum, electron-hole interactions have little effect on the optical spectrum
of isolated polyacetylene chains.
The absorption spectrum of the isolated chain obtained via
the Bethe-Salpeter equation agrees quite well with the experimental
spectrum observed for bulk $trans$-polyacetylene. However, in the previous
subsection we showed that packing effects in the bulk reduce the fundamental
band gap by about 0.5 eV. Therefore it remains to be seen whether 
Bethe-Salpeter calculations for bulk $trans$-polyacetylene would 
yield an absorption spectrum in agreement with experiment or a spectrum
shifted to lower energy.

\begin{figure}[h]
\includegraphics*[width=6.5cm]{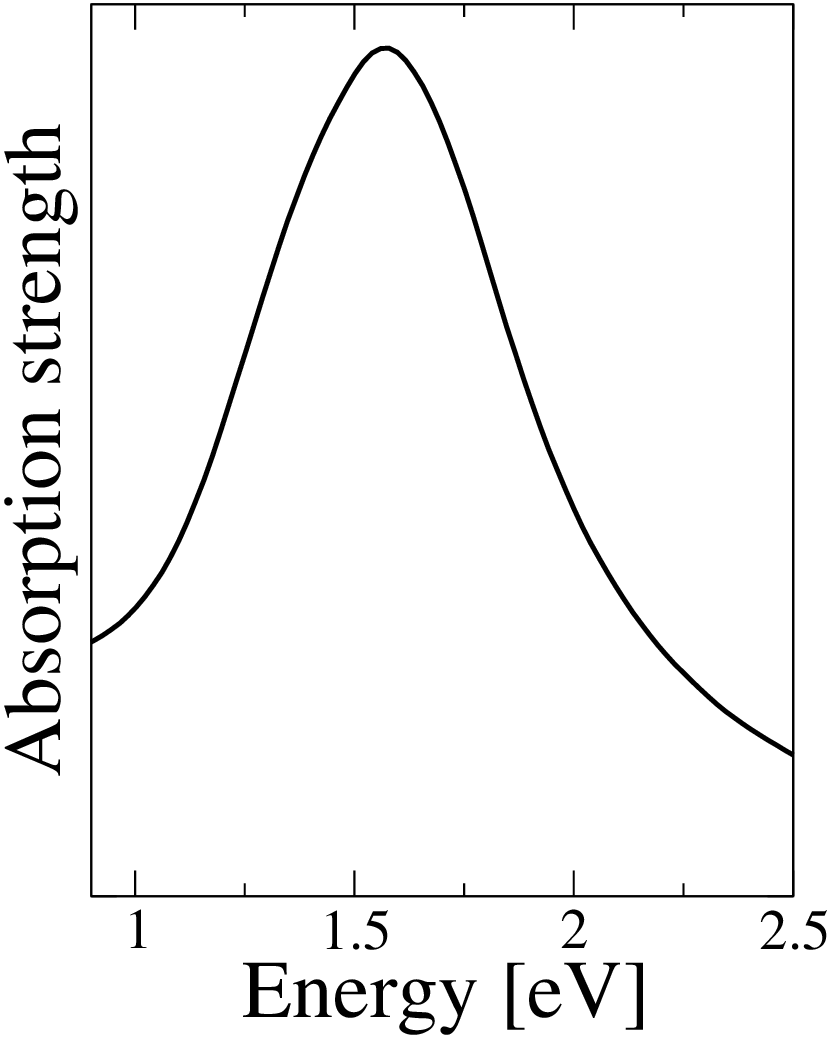}
\caption
{EXX Kohn-Sham response function for bulk polyacetylene}
\label{bulkexxindep}
\end{figure}

The optical absorption spectrum resulting from the EXX independent
particle response function 
for bulk $trans$-polyacetylene is displayed in Fig.\ \ref{bulkexxindep}.
With a value of about 1.6 eV, the maximum of the response function lies
close to the calculated values of the two band gaps at the $X$-point
and on the path between the $X$- and the $\Gamma$-point
(see Table \ref{bulkbandgaps}). Compared to the Kohn-Sham response
function for the case of an isolated, infinite chain of
$trans$-polyacetylene, the displayed result for the bulk shows
a broader peak. This widened shape, as well as the maximum position
at about 1.6 eV is in agreement with the experimental data 
for bulk $trans$-polyacetylene \cite{fincher79,leising88}.



\section{\label{sec:conclusions} Conclusions}

The exact exchange KS approach, implemented within a plane wave 
pseudopotential framework, was shown to yield band gaps for a typical 
one-dimensional periodic organic polymere,
which agree well with experimental data and with data obtained from GW methods
and methods based on the Bethe-Salpeter equation. 
Similar to the case of three-dimensional periodic semiconductors, 
EXX band gaps are substantially larger than their LDA or GGA 
counterparts and thus closer to experiment. 
The EXX approach therefore promises to be a viable tool for the 
investigation of the electronic structure of organic polymeres, a
class of materials with high potential as active compound in new
optoelectronic devices. 

The findings of this work suggest that also for polyacetylene EXX
band gaps are close to experimental band gaps, 
although in this case no direct measurements of fundamental band gaps
seem to be available and only comparisons between the EXX band gap and the
optical band gap of bulk $trans$-polyacetylene can be made.
EXX band gaps differ from the exact gaps only by 
the derivative discontinuity of the exchange-correlation potential
and the contribution of the correlation potential.
As in the case of semiconductors, the inclusion of an LDA or GGA 
correlation potential has little impact on the bandstructure, which 
suggests that quite generally the effect of the correlation potential 
can be considered to be negligible. 
Taken this for granted and assuming that fundamental and optical band gap of
bulk $trans$-polyacetylene are not grossly different, the agreement of EXX and
experimental band gaps implies that also the derivative discontinuity of the 
exchange-correlation potential is small. 
However, this point needs further investigation, because it could 
be that all presently available approximate correlation potentials 
are missing features that affect the bandstructure.

An important outcome of the present investigation is the fact that 
packing effects in bulk $trans$-polyacetylene have a substantial 
quantitative impact on the bandstructure:
The band gap in bulk $trans$-polyacetylene is about 0.5 eV smaller 
than the one obtained for isolated chains.
In future investigations of organic polymeres such
packing effects therefore should be taken into account.

\acknowledgments
We thank U. Birkenheuer and M. Rohlfing for providing the detailed 
geometry data underlying their calculations. This work was supported 
by the Deutsche Forschungsgemeinschaft. 
Calculations were carried out at the John-von-Neumann Institute of Computing. 
%

\section{\label{sec:appendix} Appendix}

In Tables \ref{geometry.bulk}, \ref{geometry.g1}, and \ref{geometry.g2}
displayed in this Appendix the geometric data for all considered systems is 
compiled. Positions (p,q,r) within the Brillouin zone are given in multiples 
$p {\bf b}_1 + q {\bf b}_2 + r {\bf b}_3$ of reciprocal lattice vectors  
${\bf b}_1$, ${\bf b}_2$, and ${\bf b}_3$ within the main text. 
The reciprocal lattice vectors 
${\bf b}_1$, ${\bf b}_2$, and ${\bf b}_3$ correspond to the lattice vectors 
${\bf a}_1$, ${\bf a}_2$, and ${\bf a}_3$, respectively. The vectors ${\bf a}_2$
and ${\bf b}_2$ are associated with the direction of the polyacetylene chains.

{\renewcommand\baselinestretch{1.10}

\begin{table}[h]
\caption{Bandgaps (in eV) for an isolated, infinite chain of $trans$-polyacetylene
calculated with various density functionals, the EXX case has been 
calculated for the two different geometries G1 and G2.
For comparison, values of Hartree-Fock (HF), M$\o$ller-Plesset theory (MP2), 
multireference configuration
interaction (MRCI) and quasiparticle calculations (GW) are given as well. \\}
\begin{tabular}{p{4cm} c} \hline\hline
& $X$ \\ \hline 
HF            &  6.06$^a$             \\  
MP2           &  3.68$^a$, 3.96$^b$   \\  
MRCI          &  4.11$^c$             \\
GW            &  2.1$^d$              \\
LDA(G1)       &  0.78                 \\  
EXX(G1)       &  1.62                 \\  
EXX(G2)       &  1.56                 \\  
EXX-VWN(G1)   &  1.65                 \\  
EXX-PBE(G1)   &  1.61                 \\  
\hline\hline \\
\end{tabular}
\label{bandgaps}
\\
\hspace{47mm} $^a$ Ref. \onlinecite{ayala01},
$^b$ Ref. \onlinecite{sun96},
$^c$ Ref. \onlinecite{birkenheuer04},
$^d$ Ref. \onlinecite{rohlfing99}\hfill \mbox{}
\end{table}

\begin{table}[h]
\caption{Bandgaps (in eV) for bulk-$trans$-polyacetylene. Beside the value for the 
gap at the $X$-point of the Brillouin zone,
a smaller gap between the $\Gamma$- and the 
$X$-point ($X\rightarrow\Gamma$) has been observed. The smallest gap was found
at the point $(0.5,0.5,0.0)$ (in units of the reciprocal lattice vectors)
lying on the edge of the Brillouin zone. The experimental values of the band
gap correspond to the energy, where a sharp rise in the absorption spectrum 
is observed, i.e., to the optical gap.\\}
\begin{tabular}{p{4cm} p{2.5cm} p{3cm} c} \hline\hline
& \centering{$\Delta_X$} & \centering{$\Delta_{X\rightarrow\Gamma}$} & 
$\Delta_{min}$  \\ \hline
LDA         & \centering{0.98} & \centering{0.81} & 0.43 \\
EXX         & \centering{1.64} & \centering{1.55} & 1.18 \\
EXX-VWN     & \centering{1.65} & \centering{1.57} & 1.21 \\
EXX-PBE     & \centering{1.61} & \centering{1.52} & 1.17 \\
Experiment  & \centering{---} & \centering{---} & 1.4$^a$, 1.5$^b$ \\
\hline\hline
\end{tabular}
\label{bulkbandgaps}
\\
\vspace{5mm}
\hspace{23mm} $^a$ Ref. \onlinecite{fincher79},
$^b$ Ref. \onlinecite{leising88} \hfill \mbox{}
\end{table}
}

\begin{table}[h]
\caption[Geometric data for bulk $trans$-polyacetylene]%
{Geometric data for bulk $trans$-polyacetylene given
in atomic units. Realspace unit cell vectors are denoted as ${\bf a_i}$,
coordinates of hydrogen (carbon) atoms are denoted as H$_i$ (C$_i$).}
\label{geometry.bulk}
\begin{center}
\begin{tabular}{p{3cm} p{3cm} p{4cm} c}
\hline\hline
& \centering{$x$} & \centering{$y$} & $z$  \\ \hline
${\bf a_1}$   & \centering{8.01244060} & \centering{0.0} & 0.0 \\
${\bf a_2}$   & \centering{-0.12144217} & \centering{4.63768905} & 0.0 \\
${\bf a_3}$   & \centering{0.0} & \centering{0.0} & 13.83279947 \\
H$_1$         & \centering{-1.76389103} & \centering{-1.153571655} & -5.367986055 \\
H$_2$         & \centering{-2.18160819} & \centering{-1.165272865} & 1.548413675 \\
H$_3$         & \centering{2.18160821} & \centering{1.165272865} & -1.548413665 \\
H$_4$         & \centering{1.88533319} & \centering{-3.484117395} & 5.367986055 \\
C$_1$         & \centering{-3.44604796} & \centering{-1.194232545} & -6.546186695 \\
C$_2$         & \centering{-0.49945126} & \centering{-1.124611975} & 0.370213055 \\
C$_3$         & \centering{0.49945126} & \centering{1.124611975} & -0.370213055 \\
C$_4$         & \centering{3.56749012} & \centering{-3.443456505} & 6.546186695 \\
\hline\hline
\end{tabular}
\end{center}
\end{table}

\begin{table}[h]
\caption[Geometric data for $trans$-polyacetylene chain (geometry G1)]%
{Geometric data for isolated chain of $trans$-polyacetylene (geometry G1) given
in atomic units. Realspace unit cell vectors are denoted as ${\bf a_i}$,
coordinates of hydrogen (carbon) atoms are denoted as H$_i$ (C$_i$).}
\label{geometry.g1}
\begin{center}
\begin{tabular}{p{3cm} p{3cm} p{4cm} c}
\hline\hline
& \centering{$x$} & \centering{$y$} & $z$  \\ \hline
${\bf a_1}$   & \centering{15.11781328} & \centering{0.0} & 0.0 \\
${\bf a_2}$   & \centering{0.0} & \centering{4.63927896} & 0.0 \\
${\bf a_3}$   & \centering{7.55890664} & \centering{0.0} & 17.00753994 \\
H$_1$         & \centering{-2.69959124} & \centering{1.10770723} & 0.0 \\
H$_2$         & \centering{2.69959124} & \centering{-1.10770723} & 0.0 \\
C$_1$         & \centering{-0.64546122} & \centering{1.11114406} & 0.0 \\
C$_2$         & \centering{0.64546122} & \centering{-1.11114406} & 0.0 \\
\hline\hline
\end{tabular}
\end{center}
\end{table}

\begin{table}[h]
\caption[Geometric data for $trans$-polyacetylene chain (geometry G2)]%
{Geometric data for isolated chain of $trans$-polyacetylene (geometry G2) given
in atomic units. Realspace unit cell vectors are denoted as ${\bf a_i}$,
coordinates of hydrogen (carbon) atoms are denoted as H$_i$ (C$_i$).}
\label{geometry.g2}
\begin{center}
\begin{tabular}{p{3cm} p{3cm} p{4cm} c}
\hline\hline
& \centering{$x$} & \centering{$y$} & $z$  \\ \hline
${\bf a_1}$   & \centering{15.11781328} & \centering{0.0} & 0.0 \\
${\bf a_2}$   & \centering{0.0} & \centering{4.67329404} & 0.0 \\
${\bf a_3}$   & \centering{7.55890664} & \centering{0.0} & 17.00753994 \\
H$_1$         & \centering{-2.69868711} & \centering{1.24389067} & 0.0 \\
H$_2$         & \centering{2.69850018} & \centering{-1.24389067} & 0.0 \\
C$_1$         & \centering{-0.61981899} & \centering{1.21122435} & 0.0 \\
C$_2$         & \centering{0.61981899} & \centering{-1.21122435} & 0.0 \\
\hline\hline
\end{tabular}
\end{center}
\end{table}

\begin{thebibliography}{40}

%
%
\bibitem{yang89} R. G. Parr and W. Yang, 
\emph{Density-Functional Theory of Atoms and Molecules} 
                 (Oxford University Press, Oxford, 1989).
%
\bibitem{dreizler90} R. M. Dreizler and E. K. U. Gross, 
                  \emph{Density Functional Theory} 
                  (Springer, Heidelberg, 1990).
%
\bibitem{koch00} W. Koch and M. C. Holthausen, 
{\em A Chemist's Guide to Density Functional Theory} 
                  (Wiley-VCH, New York, 2000).

%
%
\bibitem{hedin65} L. Hedin, Phys. Rev. {\bf 139}, A796 (1965).

%
\bibitem{hybertsen85} M. S. Hybertsen and S. G. Louie, 
                  Phys. Rev. Lett.  {\bf 55}, 1418 (1985).
%
\bibitem{aryasetiawan98} F. Aryasetiawan and O. Gunnarsson, 
                  Rep. Prog. Phys. {\bf 61}, 237 (1998).
%
\bibitem{aulbur00} W. G. Aulbur, L. J\"onsson, and J. W. Wilkins, 
                  Solid State Phys.: Adv. Res. Appl. {\bf 54}, 1 (2000).

%
\bibitem{onida02} G. Onida, L. Reining, and A. Rubio, 
                  Rev. Mod. Phys.  {\bf 74}, 601 (2002).

\bibitem{schoene98} W. D. Sch\"one and A. G. Eguiluz,
                    Phys. Rev. Lett. {\bf 81}, 1662 (1998).
%
%
%
\bibitem{hanke79} W. Hanke and L. J. Sham, 
                  Phys. Rev. Lett. {\bf 43}, 387 (1979).

%
\bibitem{albrecht98} S. Albrecht, L. Reining, R. DelSole, and G. Onida, 
                  Phys. Rev. Lett. {\bf 80}, 4510 (1998).

%
\bibitem{benedict98} L. X. Benedict, E. L. Shirley, and R. B. Bohn 
                  Phys. Rev. Lett. {\bf 80}, 4514 (1998).
%
\bibitem{rohlfing98} M. Rohlfing and S. G. Louie 
                  Phys. Rev. Lett. {\bf 81}, 2312 (1998).


%
%
%
%
\bibitem{runge84} E. Runge and E. K. U. Gross, 
                  Phys. Rev. Lett. {\bf 52}, 997 (1984).

%
\bibitem{gross96} E. K. U. Gross, J. F. Dobson, and M. Petersilka, in
                  {\it Density Functional Theory II}, 
                   Springer Series in Topics in Current Chemistry, 
                   Band 181, ed. R.F. Nalewajski 
                   (Springer, Heidelberg, 1996), p. 81.

%
\bibitem{casida95} M. E. Casida, in 
{\it Recent Advances in Density Functional Methods}, 
                  Band I, eds. D.P. Chong 
                  (World Scientific, Singapore, 1995), p. 155.

%
\bibitem{gorling98b} A. G\"orling, Int. J. Quantum Chem. {\bf 69}, 265 (1998).

%
\bibitem{gorling99d} A. G\"orling, H. H. Heinze, S. P. Ruzankin, 
                     M. Staufer, and N. R\"osch, 
                     J. Chem. Phys. {\bf 110}, 2785 (1999).          
%
\bibitem{kim02} Y.-H. Kim and A. G\"orling,
                 Phys. Rev. Lett. {\bf 89}, 096402 (2002).

%
\bibitem{berger05} J. A. Berger, P. L. de Boeij, and R. van Leeuwen,
                   Phys. Rev. B {\bf 71}, 155104 (2005).

%
%
\bibitem{perdew83} J. P. Perdew and M. Levy,
                     Phys. Rev. Lett. {\bf 51} 1884 (1983). 
%
\bibitem{sham83} L. J. Sham and M. Schl\"uter,
                  Phys. Rev. Lett. {\bf 51} 1888 (1983).


%
%
 
\bibitem{filippi94} C. Filippi, D. J. Singh and C. J. Umrigar, 
                    Phys. Rev. B {\bf 50},  14947 (1994).

\bibitem{ortiz92} G. Ortiz, 
                    Phys. Rev. B {\bf 45},  11328 (1992).

\bibitem{sham85} L. J. Sham and M. Schl\"uter, 
                    Phys. Rev. B {\bf 32},  3883 (1985).

\bibitem{perdew86} J. P. Perdew, 
                    Int. J. Quantum Chem., Symp. {\bf 19},  497 (1986).


%
%
\bibitem{perdew82} J. P. Perdew, R. G. Parr, M. Levy, and J. L. Balduz, 
                    Phys. Rev. Lett. {\bf 49} 1691 (1982).
%
\bibitem{almbladh85} C. O. Almbladh and U. von Barth, 
                      Phys. Rev. B {\bf 31} 3231 (1985).
%
%
\bibitem{chong02} D. P. Chong, O. V. Gritsenko and E. J. Baerends, 
                      J. Chem. Phys. {\bf 116} 1760 (2002).
%
\bibitem{gritsenko02} O. V. Gritsenko and E. J. Baerends, 
                         J. Chem. Phys. {\bf 117} 9154 (2002).
%
%
%
\bibitem{goerling96}  A. G\"orling, Phys. Rev. A {\bf 54} 3912 (1996).
%
\bibitem{filippi97} C. Filippi, C.J. Umrigar, and X. Gonze, 
                        J. Chem. Phys. {\bf 107} 9994 (1997).

%
%
%
\bibitem{goerling95} A. G\"orling and M. Levy, 
                        Phys. Rev. A {\bf 52} 4493 (1995).

%

%
\bibitem{kotani95} T. Kotani, 
                     Phys. Rev. Lett. {\bf 74}, 2989 (1995).
%
\bibitem{gorling96} A. G\"orling, Phys. Rev. B {\bf 53}, 7024 (1996); 
                          Phys. Rev. B {\bf 59}, 10370(E) (1999).
%
\bibitem{stadele97} M. St\"adele, J. A. Majewski, P. Vogl, and A. G\"orling, 
                     Phys. Rev. Lett. {\bf 79}, 2089 (1997).
%
\bibitem{stadele99} M. St\"adele, M. Moukara,
                     J. A. Majewski, P. Vogl, and A. G\"orling,
                     Phys. Rev. B {\bf 59}, 10031 (1999).

%
\bibitem{qteish05}  A. Qteish, A.I. Al-Sharif, M. Fuchs, M. Scheffler, 
                   S. Boeck, and J. Neugebauer, 
                   Comp. Phys. Comm.  {\bf 169}, 28 (2005).
%
\bibitem{rinke05}  P. Rinke, A. Qteish, J. Neugebauer, C. Freysoldt, 
                   and M. Scheffler,  New J. Phys.  {\bf 7}, 126 (2005).

%
\bibitem{sharma05}  S. Sharma, J. K. Dewhurst, and C. Ambrosh-Draxl, submitted.

%
%
\bibitem{aulbur00b} W. G. Aulbur, M. St\"adele and A. G\"orling,
                    Phys. Rev. B {\bf 62}, 7121 (2000).

\bibitem{fleszar01} A. Fleszar,
                    Phys. Rev. B {\bf 64}, 245204 (2001).


%
\bibitem{magyar04}  R. J. Magyar, A. Fleszar, and E.K.U. Gross, 
                     Phys. Rev. B {\bf 69}, 045111 (2004).

%
\bibitem{talman76} J. D. Talman and W. F. Shadwick, 
                   Phys. Rev. A {\bf 14}, 36 (1976).


%
%
%
\bibitem{gorling99} A. G\"orling,
                    Phys. Rev. Lett. {\bf 83}, 5459 (1999).

\bibitem{ivanov99} S. Ivanov, S. Hirata and R. J. Bartlett,
                    Phys. Rev. Lett. {\bf 83}, 5455 (1999).




\bibitem{fincherprl} C. R. Fincher Jr., C. E. Chen, A.J. Heeger ,
                     A. G. MacDiarmid, and J.B. Hastings, 
                     Phys. Rev. Lett. {\bf 48}, 100 (1982)


\bibitem{kahlert} H. Kahlert, O. Leitner, and G. Leising, 
                  Synth. Met. {\bf 17}, 467 (1987)
%





\bibitem{liegener88} C.-M. Liegener, J. Chem. Phys. {\bf 88}, 6999 (1988)

\bibitem{suhai92} S. Suhai, Int. J. Quant. Chem. {\bf 42}, 193 (1992)

\bibitem{sun96} J. Q. Sun, and R. J. Bartlett, 
                   J. Chem. Phys. {\bf 104}, 8553 (1996)

\bibitem{foerner97} W. F\"orner, R. Knab, J. Cizek, and J. Ladik,
                    J. Chem. Phys. {\bf 106}, 10248 (1997)

\bibitem{rohlfing99} M. Rohlfing, and S. G. Louie, 
                   Phys. Rev. Lett. {\bf 82}, 1959 (1999)

\bibitem{ayala01} P. Y. Ayala, K. N. Kudin, and G. E. Scuseria, 
                   J. Chem. Phys. {\bf 115}, 9698 (2001)

\bibitem{birkenheuer04} V. Bezugly, and U. Birkenheuer, 
                      Chem. Phys. Lett. {\bf 399}, 57 (2004)

%
%

\bibitem{fincher79} C. R. Fincher Jr., M. Ozaki, M. Tanaka, D. Peebles,
                     L. Lauchlan, A.J. Heeger, and A. G. MacDiarmid, 
                     Phys. Rev. B {\bf 20}, 1589 (1979)

\bibitem{leising88} G. Leising, Phys. Rev. B {\bf 38}, 10313 (1988)



%
         
%
\bibitem{sharp53} R. T. Sharp and G. K. Horton, 
                  Phys. Rev. {\bf 90}, 317 (1953).

%
\bibitem{KLI90a}                        
J. B. Krieger, Y. Li and G. J. Iafrate, 
Phys. Lett. {\bf 146} A, 256 (1990).    
%
\bibitem{EV93a}                         
E. Engel and S. H. Vosko,               
Phys. Rev. A {\bf 47}, 2800 (1993).
%
\bibitem{gorling94} A. G\"orling and M. Levy, 
                    Phys. Rev. A {\bf 50},  196 (1994).
%
\bibitem{gorling95b} A. G\"orling and M. Levy, 
                    Int. J. Quantum Chem. Symp. {\bf 29}, 93 (1995).




%
%
%
\bibitem{monkhorst76} H. J. Monkhorst and J. D. Pack,
                    Phys. Rev. B {\bf 13}, 5188 (1976).


\bibitem{froyen89} S. Froyen,
                    Phys. Rev. B {\bf 39}, 3168 (1989).



%
%
\bibitem{BK95}
D.\  M.\  Bylander and L.\  Kleinman,
Phys.\  Rev.\  Lett.\  {\bf 74}, 3660 (1995);
Phys.\  Rev.\  B {\bf 52}, 14566 (1995);
Phys.\  Rev.\  B {\bf 54}, 7891 (1996);
Phys.\  Rev.\  B {\bf 55}, 9432 (1997).


\bibitem{moukara00}
M.\  Moukara, M.\  St\"adele, J.\  A.\  Majewski, P.\  Vogl,
and A.\  G\"orling,
J.\  Phys.\  C  {\bf 12}, 6783 (2000).

\bibitem{HSC79}
D. R. Hamann, M. Schl\"uter, and C.\  Chiang,
Phys.\  Rev.\  Lett.\  {\bf 43}, 1494 (1979);
G. B. Bachelet, D. R. Hamann, and M. Schl\"uter,
Phys. Rev. B {\bf 26}, 4199 (1982).
%
\bibitem{H89}
D.\  R.\  Hamann,
Phys.\  Rev.\  B {\bf 40}, 2980 (1989).
%
\bibitem{TM91}
N. Troullier and J. L. Martins,
Phys. Rev. B {\bf 43}, 1993 (1991).

\bibitem{EHSDC01}
E.\  Engel, A.\  H\"ock, R.\  N.\  Schmid, R.\  M.\  Dreizler, 
and N.\  Chetty,
Phys. Rev. B {\bf 64}, 125111 (2001).


\bibitem{yannoni} C. S. Yannoni, and T. C. Clarke, Phys. Rev. Lett. {\bf 51}, 
1191 (1983)


\end{thebibliography}
\end{document}